\documentstyle[12pt]{article}
\newcommand{\m}{\medbreak}
\newcommand{\r}{\rightarrow}
\newcommand{\EQ}{\begin{equation}}
\newcommand{\eq}{\end{equation}}

\newcommand{\ALLPV}{\mbox{$A_{LL}^{PV}\ $}}
\newcommand{\AL}{\mbox{$A_{L}\ $}}

\begin{document}
\begin{titlepage}
\vspace{0.2in}
\vspace*{1.5cm}
\begin{center}
{\large \bf Discovery limits for a new contact interaction \\
at future hadronic colliders with polarized beams\\}
\vspace*{0.8cm}
{\bf P. Taxil} and {\bf J.M. Virey}  \\ \vspace*{1cm}
Centre de Physique Th\'eorique, C.N.R.S. - Luminy,
Case 907\\
F-13288 Marseille Cedex 9, France\\ \vspace*{0.2cm}
and \\ \vspace*{0.2cm}
Universit\'e de Provence, Marseille, France\\
\vspace*{1.8cm}
{\bf Abstract \\}
\end{center}
The production of high-transverse energy jets in hadron-hadron
collisions is sensitive
to the presence of new contact interactions between quarks.
If proton polarization were available, the
measurement of some parity violating spin asymmetries in
one-jet production at large transverse energy would complement the
usual search for deviations from the expected QCD cross section.
In the same time, a unique information on the chirality structure of
the new interaction could be obtained.
In this context, we compare the potentialities of various
$pp$ and $p\bar p$ colliders that are planned or have been proposed,
with the additional requirement of beam polarization.

\vspace*{1.0cm}

\vfill
\begin{flushleft}
\medbreak
email : Taxil@cpt.univ-mrs.fr ; Virey@cpt.univ-mrs.fr
\medbreak
CPT-96/P.3364\\
July 1996
\end{flushleft}
\end{titlepage}

\indent
As is well known,  the presence of a quark substructure can appear
in hadronic collisions as an enhancement of the one-jet inclusive
cross section at high transverse energy $E_T$.
Following Eichten et al. \cite{EichtenEHLQ} this effect
is conventionally parametrized in terms of a color singlet and
isoscalar contact term under the form :
\EQ\label{Lcontact}
{\cal L}_{qqqq} = \epsilon \, {g^2\over {8 \Lambda^2}}
\, \bar \Psi \gamma_\mu (1 - \eta \gamma_5) \Psi . \bar \Psi
\gamma^\mu (1 - \eta \gamma_5) \Psi
\eq
\noindent
where $\Psi$ is a quark doublet, $\epsilon$ is a sign and $\eta$
can take the values $\pm 1$ or 0. $g$ is a new strong coupling
constant usually normalized to $g^2 = 4\pi\,$ and $\Lambda$ is
the compositeness scale.  We assume here
that only quarks are composite, gauge bosons remaining elementary,
and also that
$\Lambda$ is much greater than the accessible subprocess energy.

Such an enhancement has been intensively searched for, in particular
at $p\bar p$ colliders \cite{UA2CDF}.  Recently, the CDF
collaboration at the Fermilab Tevatron has reported an excess of
jets at large $E_T$
with respect to the QCD prediction \cite{CDFa}. This would
correspond to a compositeness scale $\Lambda \approx 1.6$ TeV.
Although these  anomalies have to be
confirmed, they have triggered various speculations about the
possible presence of new physics effects which could be within the
reach of forthcoming  experiments \cite{AplusC}.
\medbreak \noindent
This paper is motivated by the following arguments.

As long as $\eta \neq 0$ the effective interaction
eq.(\ref{Lcontact}) violates parity
and this peculiarity should be exploited, in particular because
parity violating
effects are strictly absent in QCD.  We have found recently
\cite{TV1}  that, provided high-intensity
polarized proton beams $\vec p$ were available,
the measurement of some parity violating (PV) spin asymmetries in
one-jet inclusive production could contribute significantly to the
search  for compositeness.  Our study was performed in the context
of the Brookhaven Relativistic Heavy Ion Collider (RHIC) :
this machine will be used within a few years  by the RHIC Spin
Collaboration (RSC) as a polarized $pp$ collider \cite{RSC,BuncePw},
at a center of mass energy $\sqrt s= 500\,$GeV and with a high
luminosity  ${\cal L} = 2. 10^{32} \, cm^{-2}.s^{-1}$. With these
figures and a  degree of beam polarization ${\cal P} = 0.7$ for
each beam, spin asymmetries as  small as 1\% should be measurable in
a few months of running.

The technical progresses in the
acceleration and storage of polarized proton beams have been
impressive. They should make the same kind of measurements feasible
at machines  with higher energy, at a cost remaining a small fraction
of the cost of the collider  itself
\cite{Penn}. Various aspects of spin physics at such facilities have
been  already explored in details (see \cite{Penn,BRST,Rivista} and
references therein).

Our goal is to compare the discovery potential
and the ``analyzing power`` of some future hadronic colliders
with at least one polarized proton beam.
We will consider the RHIC collider ($\sqrt s$ = 0.5 TeV), the
$p\bar p$ Tevatron ($\sqrt s$ = 2 TeV), the upgraded
Di-Tevatron ($\sqrt s$ = 4 TeV) in the $pp$ or $p\bar p$ mode, and
the CERN LHC ($\sqrt s$ = 14 TeV), with
various possibilities for the integrated luminosity in each case.
Then, we will focus on the maximum value of the quark compositeness
scale $\Lambda$ which can be probed when parity is maximally
violated ($\eta = \pm 1$) in the  effective contact interaction
eq.(\ref{Lcontact}).

\m
For an inclusive process like $H_a\ H_b \ \r c + X$, where
$c$ is either a jet or a well-defined particle, one can define
a single-helicity PV asymmetry \AL (sometimes called the
``left-right`` asymmetry)
if only one initial hadron, say hadron $H_a$, is polarized :
\EQ
\label{ALdef}
A_{L} \;= \; {d\sigma_{a(-)b}-d\sigma_{a(+)b}\over
d\sigma_{a(-)b}+d\sigma_{a(+)b}}
\eq
\noindent
where the signs $\pm$ refer to the helicities of the colliding
hadrons. This quantity \AL is the only relevant one in case of
$\vec p \, \bar p$ collisions since there is no known way to get
intense and highly energetic  polarized antiproton beams.
When both proton beams can be polarized (this is the case at RHIC),
one  defines a double helicity PV asymmetry :
\EQ
\label{ALLPVdef}
A_{LL}^{PV} ={d\sigma_{a(-)b(-)}-d\sigma_{a(+)b(+)}\over
d\sigma_{a(-)b(-)}+d\sigma_{a(+)b(+)}}
\eq
From now, $d\sigma_{a(h_a)b(h_b)}$ will mean the
cross section in a given helicity configuration $(h_a,h_b)$,
for the production of a single jet
at a given transverse energy $E_T$ and pseudorapidity $\eta$ :
\EQ
\label{dsrap0}
d\sigma_{a(h_a)b(h_b)}
\; \equiv \; {d^2\sigma^{(h_a)(h_b)} \over {dE_T d\eta}}
\eq
\noindent
In the following, we choose to integrate $d\sigma$ over a
pseudorapidity interval
$\Delta \eta =1$ centered at $\eta=0$, and over an $E_T$ bin which
corresponds to a jet energy resolution of 10\% \cite{UA2CDF}.
\medbreak
Any helicity dependent hadronic cross section is obtained by
convoluting appropriately
the subprocess cross sections
${d\hat \sigma}_{ij}^{\lambda_1,\lambda_2}/d {\hat t}\;$, which
depend upon the
parton helicities $\lambda_1$ and $\lambda_2$, with the polarized
quark and/or antiquark distributions  evaluated at some scale $Q^2$:
$q_{i\pm}(x,Q^2)$ and $\bar q_{i\pm}(x,Q^2)$ (explicit formulas can
be found in \cite{BRST,Rivista,BouGuiSof}).
Here, $q_{i\pm}$ means the distribution of the polarized quark of
flavor $i$ having its helicity parallel (+) or antiparallel (-) to
the parent hadron  helicity. The chosen $Q^2$ value is
$Q^2 = E_T^2$, we have checked that  changing this choice has
no visible influence on our results.
In the following $\hat s$, $ \hat t$ and $\hat u$ denote the usual
Mandelstam variables for the subprocess $i_a\ j_b \ \r k + l$
and are given, at zero rapidity, by :
\EQ
\hat s = x_a\,x_b\,s,\;\;\;\;\; \hat t = -\,x_a\,x_T\,s/2,\;\;\;\;\;
\hat u = -\,x_b\,x_T\,s/2 
\eq
\noindent with $x_T \equiv 2E_T/{\sqrt s}$.\\
Concerning the subprocess cross sections we follow the notations
of \cite{BouGuiSof} where :
\EQ
{d\hat \sigma_{ij}^{\lambda_1,\lambda_2}\over d\hat t}
\; =\;{\pi\over \hat s^2} \,
\sum_{\alpha,\beta} T_{\alpha,\beta}^{\lambda_1,\lambda_2}(i,j)
\eq
\noindent
$T_{\alpha,\beta}^{\lambda_1,\lambda_2}(i,j)$ denoting the matrix
element squared with $\alpha$ boson and $\beta$ boson exchanges, or
with one exchange process replaced by a contact interaction.
These terms will be evaluated at leading order.
Note that one-loop QCD corrections for inclusive jet production in
composite modelshave been recently estimated  \cite{Lee}. QCD being
helicity conserving in the limit of massless quarks, one does not
expect a significant influence of such corrections on the spin
asymmetries.

QCD being also parity conserving, only the direct Contact term
$T_{CT.CT}$,  the one-gluon exchange-Contact interference term
$T_{g.CT}$ and the terms involving Electroweak (EW) gauge bosons
exchanges are involved in the calculations of the numerators of \AL
and \ALLPV. Of course, when evaluating the denominators in
eqs.(\ref{ALdef}), and  (\ref{ALLPVdef})
-that is the unpolarized cross
section which is QCD dominated- all the terms, involving quarks,
antiquarks and also gluons, have to be included.

The terms involving Contact amplitudes have the following expressions
\cite{Rivista} :
\medbreak
- For identical quarks $q_i q_i \r q_i q_i$ :
\EQ\label{CTCT}
T_{CT.CT}^{\lambda_1,\lambda_2}(i,i) \; =\;
{8 \over 3} \, {{\hat s}^2 \over \Lambda^4} \,
(1 - \eta \lambda_1)(1 -  \eta\lambda_2)
\eq
\noindent
the crossed process $ \, q_i  \bar q_i \r  q_i \bar q_i \, $ is
obtained by changing $\hat s \r \hat u$ and $\lambda_2 \r
-\lambda_2$. In case of scattering of identical antiquarks, change
$\lambda_1,\lambda_2$ into $-\lambda_1,-\lambda_2$ in
eq.(\ref{CTCT}). \medbreak
- For quarks of different flavors $q_i q_j \r q_i q_j\; (i\neq j)$  :
$T_{CT.CT}^{\lambda_1,\lambda_2}(i,j) \; =\; (3/8)\;
T_{CT.CT}^{\lambda_1,\lambda_2}(i,i)$,
with the same changes as above for the crossed processes
$q_i \bar q_j \r q_i \bar q_j$  as well as for $q_i \bar q_i \r q_j
\bar q_j$. For antiquarks  $\bar q_i \bar q_j \r \bar q_i \bar q_j$,
change  $\lambda_1,\lambda_2$ into $-\lambda_1,-\lambda_2$ in the
first expression. \medbreak
Due to color conservation rules, the interference between the
one-gluon exchange amplitude and the Contact term amplitude occurs
only for identical quarks (identical antiquarks), therefore for
$q_i q_i \r q_i q_i$ : \EQ\label{gCT}
T_{g.CT}^{\lambda_1,\lambda_2}(i,i) \; =\;
{8 \over 9} \alpha_s(Q^2)\, {\epsilon \over \Lambda^2} \,
(1 - \eta \lambda_1)(1 -  \eta\lambda_2)\left({\hat s^2 \over \hat t}
+ {\hat s^2 \over \hat u}\right)
\eq
\noindent
with the change $\lambda_1,\lambda_2$ into $-\lambda_1,-\lambda_2$
for $ \, \bar q_i  \bar q_i \r  \bar q_i \bar q_i \, $ or
$\hat s \r \hat u$ and $\lambda_2 \r -\lambda_2$  for
$ q_i  \bar q_i \r   q_i \bar q_i  $. Note that $\hat t$ and $\hat u$
being negative, $\epsilon=-1$ (+1) corresponds to constructive
(destructive) interference \cite{EichtenEHLQ}.
\medbreak
We have checked that the influence of the interference between
EW and CT amplitudes is quite weak (the expressions for the dominant
terms can be found in \cite{TV1}).
We will call generically ASM
the PV asymmetry (\AL or \ALLPV) which is expected in the Standard
Model as due to EW boson exchanges and also to QCD-EW interference.
These standard PV asymmetries  have been
studied for a long time \cite{AbudBaurGloverMartin,PaigeRanft} and
the correct expressions for the helicity dependent amplitudes can be
found in \cite{BouGuiSof}. In any case,  ASM remains small although
it increases in magnitude with $E_T$ at a fixed $\sqrt s$ value
\cite{TannenbaumPenn}.  This is due to the increasing importance of
quark-quark  scattering relatively to other terms involving gluons.
The non-standard asymmetries exhibit the same behavior.
For illustration we give in Table 1 the values obtained for ASM
in various collider configurations for a value of
$x_T \approx 1/3$, which is
relevant for our study.
\begin{table}[b]
\begin{center}
\begin{tabular}{|c|c|c|}
\hline
 Collider & ASM (\%) & $\Delta A$ (\%)\\
\hline
RHIC ($pp$), $ L_1=0.8 fb^{-1}$ & $1.36\pm0.6$ & 0.7\\
\hline
Tevatron ($p\bar p$), $L_1=1 fb^{-1}$ & $-0.33\pm0.12$ & 2.6\\
\hline
Di-Tevatron ($p\bar p$), $L_1=10 fb^{-1}$ & $-0.34\pm0.13$ & 2.0\\
\hline
Di-Tevatron ($pp$), $L_1=10 fb^{-1}$ & $3.11\pm1.11$ & 3.0\\
\hline
LHC ($pp$), $L_1=100 fb^{-1}$ & $4.03\pm1.0$ & 6.8\\
\hline
\end{tabular}
\end{center}
\caption{
Standard ASM (\ALLPV for $pp$, \AL for $p\bar p$)  for
$x_T \approx 1/3$, at various colliders with integrated luminosity
$L_1$ along with the statistical error $\Delta A$ on ASM.
}
\end{table}
In this table, ASM is  given in the first column  with a
``theoretical`` error which corresponds to our estimate of the
present uncertainties due to the imperfect knowledge of the
polarized quark and antiquark distributions. For this purpose we
have used some recent sets of  distributions  (GS95 \cite{GS95},
GS96 \cite{GS96}, GRV \cite{GRV} and BS \cite{BS95}) which fit all
the available data from polarized deep-inelastic experiments. Note
that, since real gluons are not involved in the process we consider,
our estimates are not plagued   by the uncertainties associated to
the imperfect knowledge of the polarized gluon distributions
$\Delta G(x,Q^2)$. On the other hand, the statistical error
$\Delta A$ for an integrated luminosity $L_I$ is
given by (for \ALLPV) :
\EQ
\Delta A\; \simeq\; {1\over {\cal P}}\, {\sqrt {L_1 \over L_I}}
{1\over {\sqrt{N_{evts}^{++}\,+\,N_{evts}^{--}}}}
\eq \noindent
where the number of events $N_{evts}^{++(--)}$
corresponds to  $L_1/4$.
When the measurement of \AL is concerned one has :
$\Delta A$(\AL) $\, \simeq {1\over \sqrt{2}} .\Delta A$(\ALLPV). In general,
\ALLPV is larger than \AL in the same kinematical conditions, for a
statistical error which is comparable. Therefore, it is better to
retain the former when its measurement is feasible.

Of course, in a given hadronic configuration ($pp$ or
$p\bar p$), when $\sqrt s$ increases at fixed $x_T$, a greater
luminosity is needed to get the same number of events, and therefore
the same value for $\Delta A$.

Concerning ASM ($\equiv$ \AL) in $p\bar p$ collisions, it is
dominated by $q \bar q$ annihilation and its magnitude is very small.
This is due, first, to the crossing symmetry : annihilation terms
contribute much less than scattering terms to the numerator of \AL;
second, to the important cancellation which occurs between the
QCD-EW interference terms ($T_{gZ}$ or $T_{gW}$) on the one hand,
and the pure EW terms which are relevant to the particular process
($T_{WW}$, $T_{ZZ}$, $T_{\gamma Z}$) on the other hand
\cite{BouGuiSof}. \medbreak Turning now to the search for
non-standard effects, we give in Fig. 1. the  95\% confidence level
limits on the compositeness scale $\Lambda$  which could be obtained
at the future colliders we consider. We compare the limits obtained
from measurements of the unpolarized one-jet cross-section alone and
from the measurement of the PV spin asymmetry \AL or \ALLPV in the
same channel. The strategies which have been followed are based on a
$\chi^2$ analysis, they are described below :\\

- {\it Cross sections} :
\\
Using GRV distributions, we have compared $d\sigma$(QCD+EW+CT) to
the standard QCD-dominated cross section. To calculate the
$\chi^2$, we have added in quadrature a systematic uncertainty of
50\% to the statistical uncertainty \cite{CDFa}. As noticed by CDF
\cite{UA2CDF}, this kind of analysis is dominated by the upper part
of the $E_T$ spectrum. As a check, we recover the published CDF
limit, $\Lambda=1.4$ TeV, obtained with a data sample of 4.2
$pb^{-1}$\cite{UA2CDF}. Using some other (unpolarized) quark
distributions which are currently in use yields the same results.

- {\it Asymmetries} :
\\
In this case, the strategy is different since a reasonable
number of events is necessary for measuring an asymmetry. Therefore,
the analysis is dominated by the region
\linebreak $x_T \approx 0.25 - 0.4$.
On the other hand, in an asymmetry which is a ratio of cross
sections, certain systematic errors such as detector efficiencies
and absolute luminosities cancel \cite{RSC}. We have been
conservative, choosing $(\delta A/A)_{syst} = 20$\%. We have also
chosen a set of polarized distributions (GRV) in which the quarks
carry a small fraction of the proton spin, at a variance with e.g.
the BS distributions. The magnitude of the spin asymmetries are then
reduced and the bounds we give on $\Lambda$ are conservative. It has
also to be kept in mind that our knowledge about the polarized
partonic distributions will improve drastically in the future,
thanks to the HERMES experiment at HERA \cite{HERMES} and,
especially, thanks to the RSC program itself (see e.g.
\cite{BuncePw,BouGuiSof}). \medbreak One can see from Fig.1 that
measuring the PV spin asymmetry gives in general much better
discovery limits for the compositeness scale than the measurement of
the unpolarized cross section. This behaviour is independent of the
left-handed or right-handed nature of the new interaction, as long
as PV is maximal. The key factor turns out to be the integrated
luminosity :  at a fixed c.m. energy, the spin asymmetry gives a
better sensitivity as soon as the large luminosity allows to get a
big number of events with $x_T \geq 0.3$. The situation at RHIC is a
particular case since the low value of $\sqrt s$ yields some bounds
which are below the present CDF limit (from the measurement of
$\sigma$). Note that the bounds we obtain from \ALLPV are larger by
$\approx 1.3$ TeV than the ones we had obtained in a simpler
analysis  \cite{TV1}.

In each case, we have made the distinction between
constructive and destructive  interference between QCD and CT
amplitudes. As already noticed in EHLQ \cite{EichtenEHLQ}, the
non-standard effect is more visible when $\epsilon = -1$. This is
true from the cross section and also from the PV spin asymmetry.
However,  the respective weights of the
direct $T_{CT.CT}$ term and the $T_{g.CT}$ interference term vary in
function of the collider ($pp$ or $p\bar p$) configuration. As a
consequence, in the $pp$ mode, the bounds on $\Lambda$ are more
influenced by the sign of $\epsilon$ when they come from the cross
section measurement than from the asymmetry. The situation is
reversed in the $p\bar p$ mode.  In any case, at a given luminosity,
the Di-Tevatron is preferred in the $pp$ mode.

Finally, it is important to note that, if an effect is observed in
jet production, the polarized collider will not only  be a tool for
discovery but also for analysis since  it will provide valuable
information on the chirality structure of the new interaction. More
precisely, from the way \AL or \ALLPV deviates from the ASM value,
it is easy to get the sign of the product $\epsilon \eta$, even for
the largest value of $\Lambda$ in a given collider configuration.
For instance, at RHIC for $x_T = 1/3$ with the GRV distributions,
the Standard Model expectation is \ALLPV (SM) = 1.3 \% (with an error
of 0.8 \% with $L_1=0.8 fb^{-1}$). Adding the contact interaction
with a scale $\Lambda = 1.6$ TeV, we obtain
\ALLPV ($\epsilon \eta = 1$) = $-$ 1.2 \% and
\ALLPV ($\epsilon \eta = -1$) = 3.9 \%.

\medbreak We thank J. Soffer for discussions and information about
the RHIC  spin physics program, L. Lellouch for comments
and C. Benchouk, C. Bourrely, M.C. Cousinou, E. Nagy, and
C. Vall\'ee for useful advice about the analysis.
J.M. Virey is Moniteur CIES and allocataire MESR.
Centre de Physique Th\'eorique is UPR 7061


\newpage  \noindent
{\bf Figure captions}
\bigbreak
\noindent
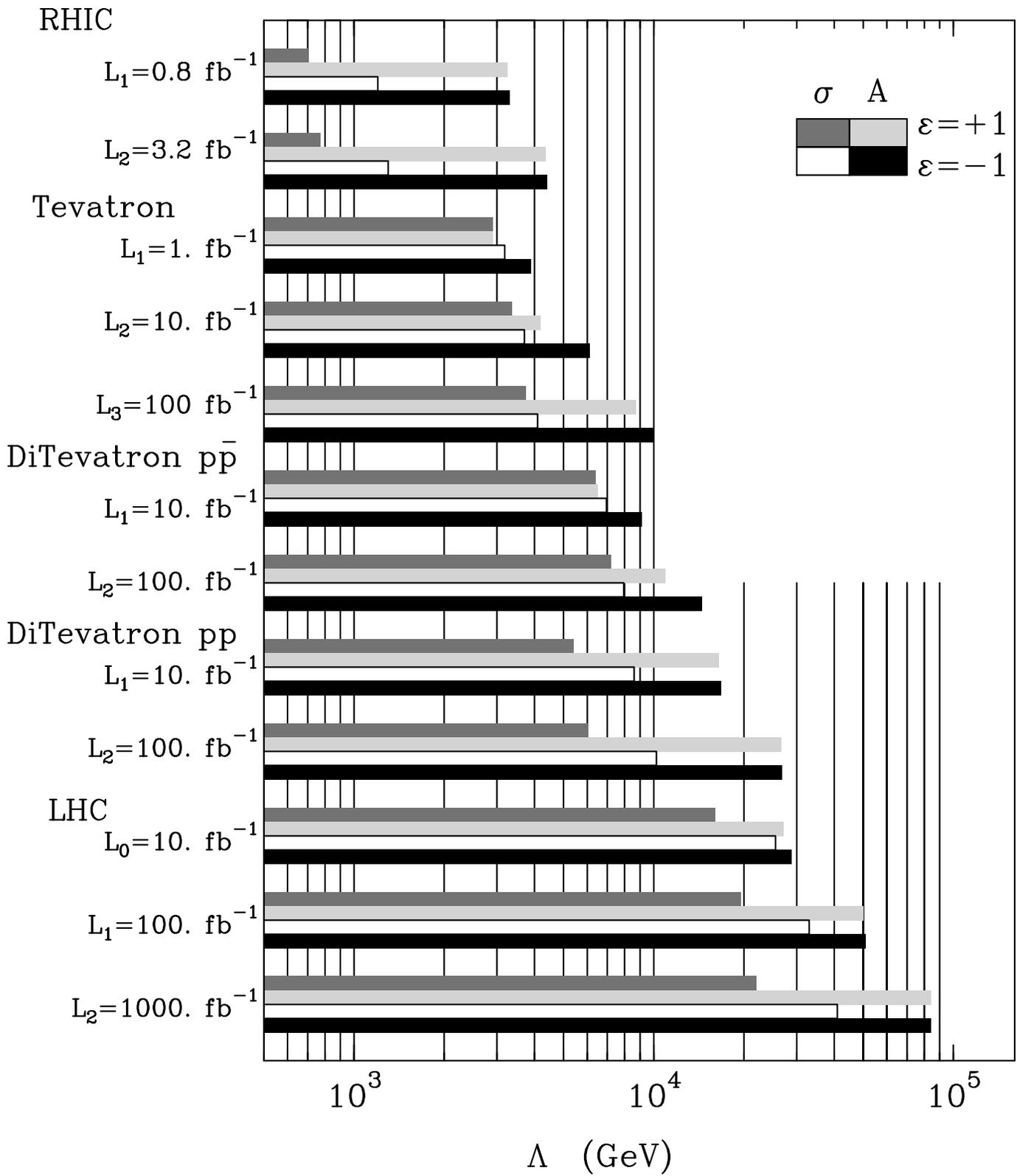
{\bf Fig. 1} 95\% C.L. discovery limits for the compositeness scale
$\Lambda$ at future hadronic colliders with polarization, in case of
constructive ($\epsilon = -1$) or destructive ($\epsilon = +1$)
interference between the QCD and the non-standard amplitudes. The
$\chi^2$ analysis is based on the unpolarized one-jet cross section
or independently on the PV spin asymmetry \ALLPV (\AL in case of
$p\bar p$ collisions). These limits are independent of the sign of
the parameter $\eta$ ($\eta = \pm 1$).


\end{document}